\documentclass{aa}
\usepackage[varg]{txfonts}

\usepackage{graphicx}
\usepackage[colorlinks=true, allcolors=blue]{hyperref}
\usepackage{enumitem}

\begin{document}

\title{Closing up the cluster tension?}

\author{
   A.~Blanchard\inst{1} \and
   S.~Ili\'c\inst{2, 1}
}

\institute{
    Universit\'e de Toulouse, UPS-OMP, IRAP and CNRS, IRAP, 14, avenue Edouard Belin, F-31400 Toulouse, France
    \and
    Universit\'e PSL, Observatoire de Paris, Sorbonne Universit\'e, CNRS, LERMA, F-75014, Paris, France \\
    \email{alain.blanchard@irap.omp.eu, stephane.ilic@obspm.fr}
}

\abstract{The excellent measurements of the cosmic microwave background (CMB) fluctuations by Planck allow us to tightly constrain the amplitude of matter fluctuations at redshift $\sim 1100$ in the $\Lambda$-cold dark matter ($\Lambda$CDM) model. This amplitude can be extrapolated to the present epoch, yielding constraints on the value of the $\sigma_8$ parameter. On the other hand, the abundance of Sunyaev-Zeldovich (SZ) clusters detected by Planck, with masses inferred using a hydrostatic equilibrium assumption,
leads to a significantly lower value of the same parameter. This discrepancy is often dubbed the $\sigma_8$ tension in the literature and is sometimes regarded as a possible sign of new physics. Here, we examine a direct determination of $\sigma_8$ at the present epoch in $\Lambda$CDM, and thereby
the cluster mass calibrations using cosmological data at low redshift, namely the measurements of $f\sigma_8$ from the analysis of the completed Sloan Digital Sky Survey (SDSS). We combined redshift-space distortion measurements with Planck CMB constraints, X-ray, and SZ cluster counts within the $\Lambda$CDM framework, but leaving the present-day amplitude of matter fluctuations as an independent parameter (i.e. no extrapolation is made from high-redshift CMB constraints). The calibration of X-ray and SZ masses are left as free parameters throughout the whole analysis. Our study yields tight constraints on the aforementioned calibrations, with values entirely consistent with results obtained from the full combination of CMB and cluster data only. Such an agreement suggests an absence of tension in the $\Lambda$CDM model between CMB-based estimates of $\sigma_8$ and constraints from low-redshift on $f\sigma_8$; however, it also indicates tension with the standard calibration of clusters masses.}


\maketitle

\section{Introduction}
\label{sec:Intro}

The measurements of the fluctuations of the cosmic microwave background (CMB) by Planck has provided a remarkable test of the $\Lambda$-cold dark matter ($\Lambda$CDM) picture. Nearly forty years after its foundation \citep{Peebles82,1984ApJ...284..439P}, the model is in tight agreement with CMB data, allowing us to estimate its main parameters to percent-level accuracy \citep{Param-Planck2018}. Data from the Wilkinson Microwave Anisotropy Probe combined with measurements from the Atacama Cosmology Telescope on small scales lead to essentially identical values with similar uncertainties \citep{2020JCAP...12..047A}, leaving little room for remaining systematics. However, estimates of several parameters obtained from low-redshift probes are strikingly different from Planck best-fit values, with relatively high significance. The disagreement on the value of the local Hubble constant has attracted a lot of attention \citep{Riess} and is currently considered as a possible indication of tension between late and early Universe physics \citep{verde}. Similarly, measurements of the amplitude of matter fluctuations at low redshift from lensing surveys yield values lower than those inferred from the CMB \citep{2021A&A...646A.140H}, although the issue remains a matter of debate. From the first year of the Dark Energy Survey (DES Y1), the $3\times 2$ pt analysis \citep{2018PhRvD..98d3526A} leads to a lower amplitude, although the collaboration concludes that the two data sets are consistent, while \citet{2021MNRAS.505.6179L} derived a 2.3\,$\sigma$ tension. However, the recent DES Y3 analysis from the $3\times2$pt data \citep{2021arXiv210513549D} concludes that their data are consistent with the predictions of the model favoured by the Planck 2018 data.

The abundance of galaxy clusters and its evolution with redshift are known to provide interesting cosmological constraints widely discussed in the literature \citep{1992A&A...262L..21O,1993MNRAS.262.1023W,1993ApJ...407L..49B,1993ApJ...407L..45B,1995A&A...300..637H,1997ApJ...489L...1H}, with a strong sensitivity to the growth rate of cosmic structures \citep{bb}. Recent analyses of those abundances provide an additional source of tension on the amplitude of matter fluctuations. Indeed, Sunyaev-Zeldovich (SZ) clusters number counts as measured by Planck have been found to be much lower than expected from the CMB-only best-fit $\Lambda$CDM model when using a fiducial mass-SZ signal calibration, derived from X-ray observations coupled to hydrostatic mass estimations \citep{2014A&A...571A..20P,2016A&A...594A..24P}.  Those results indicate a value of the amplitude of matter fluctuations more in line with the weak lensing results previously mentioned.

The aforementioned discrepancies, which arise when comparing results from low-redshift probes and CMB data, have opened the door to new physics, although no simple solution to these tensions has emerged yet \citep{Jedamzik}. Cluster data from DES Y1 yielded a surprisingly low amplitude for matter fluctuations and the density parameter $\Omega_m$ \citep{2020PhRvD.102b3509A} and is regarded as a clear sign of the presence of systematics. It is therefore of high interest to find probes of the amplitude of matter fluctuations that are independent of lensing data. The purpose of the present work is to examine the amplitude of matter fluctuations from the eBOSS redshift-space distortion  (RSD) measurements in the $\Lambda$CDM model and its implications for cluster mass calibration.

In Section~\ref{sec:MassCalib}, we summarise the calibration-related issues that appear when using cluster samples as cosmological probes. We recall the tension between the amplitude of the calibration required to fit the Planck SZ cluster counts compared to the fiducial calibration adopted in the Planck analysis of the same sample \citep{2014A&A...571A..20P}. This tension between cluster counts and CMB observations is a long-standing problem \citep{2005A&A...436..411B}, the significance of which was raised above 4\,$\sigma$ by the Planck results. In Section~\ref{sec:CalibWithSDSS}, we present the self-calibration approach we followed in order to derive constraints on $\Lambda$CDM matter amplitude from eBOSS data, using Planck priors on other cosmological parameters. In Section~\ref{sec:Results}, we compare calibrations obtained with and without the addition of the RSD measurements from the completed Sloan Digital Sky Survey (SDSS) in a variety of cosmological scenarios, and we draw our conclusions in Section~\ref{sec:Conclusion}.
\section{The mass calibration issue}
\label{sec:MassCalib}
In order to use cluster abundances as cosmological probes, in particular when using small samples, it is necessary to assume a relation (and its dispersion) between the total cluster mass and its observable.
 Such relations can be derived from scaling arguments \citep{1986MNRAS.222..323K,1991ApJ...383..104K}, and are in good agreement with results from hydrodynamical numerical simulations \citep{1998ApJ...495...80B}. The calibration of these scaling relations is, however, more uncertain.
On their own, hydrostatic mass estimations are quite imprecise \citep{1997ApJ...487...33B}.
 Furthermore, numerical simulations of clusters have shown the importance of non-gravitational heating of their gas \citep{{1998ApJ...507...16S}}. In consequence, the degeneracy between the amplitude of matter fluctuations and the cluster calibration remains when we try to extract cosmological constraints from such probes \citep{2003MNRAS.342..163P}. This is sometimes referred to as the `calibration issue' of galaxy clusters. It is therefore safer to leave the calibration as a free parameter \citep{2010A&A...524A..81D}.

In the present work, we made use of both X-ray and SZ clusters (see Section~\ref{sec:CalibWithSDSS}), thus requiring some assumptions regarding the scaling laws of the mass proxies corresponding to those samples.
We used the following relation for the temperature of X-ray clusters accordingly to the scaling law expressed at the virial mass of the cluster $M_v$:
\begin{equation}
 \label{eq:MTscaling}
T=A_{\rm {TM}}(h\, M_v)^{2/3}\left(\frac{\Omega_{m} \Delta_b(z)}{178}\right)^{1/3}(1+z)\,,
\end{equation}
where $A_{\rm {TM}}$ is the X-ray calibration constant, and $\Delta_b(z)$ is the virial density contrast with respect to the total background matter density of the Universe at redshift $z$.

Sunyaev-Zeldovich-selected samples of clusters are of strong interest for cosmology as the SZ signal is expected to be tightly related to the total cluster mass through their scaling relations \citep{1996A&A...314...13B}. Numerical simulations have shown that such mass proxy does seem reliable \citep{2006ApJ...650..128K}. For SZ-detected clusters, we used the following expression for the scaling relation between the integrated Compton $y$-profile $\bar{Y}_{500}$ and mass, in accordance with the choice of \citet{2016A&A...594A..24P}:
\begin{equation}
\label{eq:Yscaling}
E^{-\beta}(z)\left[\frac{{D_{\rm A}}^2(z) {\bar{Y}_{500}}}{\mathrm{10^{-4}\,Mpc^2}}\right] =
Y_\ast \left[ {\frac{h}{0.7}} \right]^{-2+\alpha} \left[\frac{(1-b)\,{M_{500}}}{6\times10^{14}\,M_{\odot}}\right]^{\alpha}.
\end{equation}
We used the fiducial values from \citet{2016A&A...594A..24P} for the parameters $\log Y_\ast = -0.19$ and $\beta =0.66$, and the Gaussian prior on the dispersion of the scaling law $\sigma_{\ln Y}=0.127\pm0.023$. The slope parameter $\alpha$ is left free throughout the analysis. The $M_{500}$ cluster mass, used for cluster identification in the Planck SZ sample, is a convention based on defining the density contrast of clusters 500 times the critical density of the Universe. The $b$ parameter is the so-called hydrostatic bias. The fiducial value of $(1-b)$ was taken to be $0.8$ in the Planck analysis \citep{2014A&A...571A..20P}. Its actual value is a key ingredient for the interpretation of the cluster counts. Once this calibration is set, the abundance of clusters (as a function of their mass and redshift) in a survey given its selection criteria can be computed using the mass function of halos and compared to observations. Such an abundance and its evolution are thus powerful cosmological tools \citep{bb} that are widely discussed and applied in the literature. For large samples with a well-controlled selection function, a self-calibration procedure can be used \citep{Hu2003}, avoiding the use of prior (and potentially biased) knowledge of observable-mass relation. The Planck SZ cluster sample is, however, too small to obtain stringent constraints,
and one must settle on a mass calibration to obtain useful conclusions. For this reason, a fiducial external calibration $(1-b) = 0.8$ was chosen based on early investigation of X-ray clusters and taking into account results from hydrodynamical simulations \citep{2014A&A...571A..20P}. In the Planck best-fit $\Lambda$CDM cosmology, this calibration leads to a number of predicted SZ clusters nearly three times larger than the observed counts, while a value close to 0.6 allows us to obtain an agreement.

As a preliminary result, we updated here the 
constraints on the calibration $(1-b)$ in the $\Lambda$CDM model compared to \citet{2019A&A...631A..96I} using the Planck 2018 CMB data \citep{2020A&A...641A...5P} and the mass function from \citet[][D16 hereafter]{2016MNRAS.456.2486D}. We found that the tension has slightly increased, as the 68\% confidence limits on the calibration yield $(1-b)= 0.622\pm 0.028$, more than 6\,$\sigma$ away from the fiducial 0.8 value.
 However, this Planck CMB-based calibration 
does not stem directly from a measurement of the amplitude of matter fluctuations at low redshift. Rather, it arises from amplitude estimates at redshift $\sim 1100$, extrapolated to low redshifts according to the $\Lambda$CDM model.

This cluster calibration issue limits the potential power of cluster counts to constrain cosmology \citep{2018PhRvD..97j3531P}. Most critically, a definitive piece of evidence for $(1-b)\sim 0.8$ would reveal a low-redshift amplitude of matter fluctuations inconsistent with the standard $\Lambda$CDM picture, which simple extensions such as massive neutrinos or a phenomenological $\gamma$-modification of gravity cannot explain \citep{2019A&A...631A..96I}. In the next section, we examine this calibration issue using an additional and independent source of low-redshift data. Our aim is to further examine the presence of a potential tension in the $\Lambda$CDM model between observations of the early and late Universe, without using any external prior on cluster mass calibrations.
\section{Cluster self-calibration using low-redshift SDSS data}
\label{sec:CalibWithSDSS}

The extended Baryon Oscillation Spectroscopic Survey (eBOSS) Collaboration recently presented a summary of the cosmological constraints obtained from various surveys conducted within the SDSS in a recent paper \citep{2021PhRvD.103h3533A}. Among those results, the growth rate of matter fluctuations was measured over a large redshift range via RSD (cf.~Table~3 in the aforementioned reference). RSD measurements are indeed likely to provide reliable constraints \citep{2021MNRAS.tmp..238H} on the growth of structures. More specifically, RSD provides a way of measuring $f(z)\sigma_8(z)$, where $f(z)$ is the so-called growth rate of matter:
\begin{equation}
    f(z) \equiv \frac{{\rm d}\ln \sigma_8}{{\rm d}\ln a}\,,
\end{equation}
while $\sigma_8(z)$ is a measure of the amplitude of matter fluctuations at redshift $z$. These measurements rely only on data from the clustering of galaxies in redshift space. Inferring the amplitude of $\sigma_8$ from the eBOSS RSD data alone does not lead to stringent limits (see Fig.~9 in \cite{2021PhRvD.103h3533A}). Priors on cosmological parameters help to reduce these uncertainties (see Fig.~11 in the same reference).
 Here, we want to use the eBOSS RSD data in combination with Planck CMB data in order to provide a self-calibration for clusters within the $\Lambda$CDM model, without the addition of any external prior on the mass calibration. We note that RSD and CMB data are essentially independent and can be combined easily without accounting for their covariance. In the following, we use as data sets:
 (i) the aforementioned RSD measurements from eBOSS, specifically the RSD-only line in Table~3 of \citealt{2021PhRvD.103h3533A}; (ii) the latest CMB data from Planck, including the full fiducial likelihoods for the low- and high-multipole temperature and E-mode polarisation, released in 2018; (iii) the local sample of X-ray clusters from \citet{2015A&A...582A..79I}; (iv) the sample of SZ clusters from \citet{2016A&A...594A..24P}.

 As stated earlier, our objective is to infer the galaxy cluster mass calibration with the addition of eBOSS RSD data to the Planck CMB data via the self-calibration approach. A direct combination of the two (with cluster samples) leads to tight constraints on the calibration due to the stringency of Planck data. The effect of adding eBOSS RSD data leads to constraints that are indistinguishable from Planck-only ones, as illustrated by the dashed lines and contours in Fig.~\ref{Triplot_S8}. Those results are therefore essentially identical to the Planck-only $\Lambda$CDM results of \citet{2019A&A...631A..96I} mentioned in Section~\ref{sec:MassCalib} (apart from the update to 2018 data). However, we remind the reader that the tightness of those CMB-based constraints on the cluster calibration may be misleading. It is only through the extrapolation of the tight early-Universe constraints on matter fluctuations that the CMB is able to break the known degeneracy between $(1-b)$ and (an extrapolated) $\sigma_8$.

 In order to obtain constraints on the current value of $\sigma_8$ -- and thus on the cluster calibration -- that stem primarily from the eBOSS data, we slightly tweaked the $\Lambda$CDM model
 by introducing an additional degree of freedom. More specifically, the growth factor of structures is multiplied by a constant -- left as a free parameter -- over the redshift range corresponding to our cluster samples (roughly $z\sim[0,1]$). In practice, this allows us to decouple from early times and directly control the present value of $\sigma_8$ in the model. Once we introduce data, the behaviour of the background (at all times) and the perturbation sector (up to $z\sim 1$) remain similarly constrained by the CMB. Meanwhile, $\sigma_8$ is almost entirely constrained by low-redshift probes since the CMB is mostly insensitive to late-Universe physics.

 Such a method of rescaling matter fluctuations is acceptable since in $\Lambda$CDM models (and other minimally coupled dark energy models) the late-time growth is scale independent. This property was used, for instance, in \citet{IST:paper1} (when exploring a simple modification of gravity) or for the DES Y1 analysis presented in \citet{2021PhRvD.103b3528M}.

\section{Results: Matter fluctuation amplitude and clusters calibration}
\label{sec:Results}

\begin{figure}
\sidecaption
  \includegraphics[width=\columnwidth]{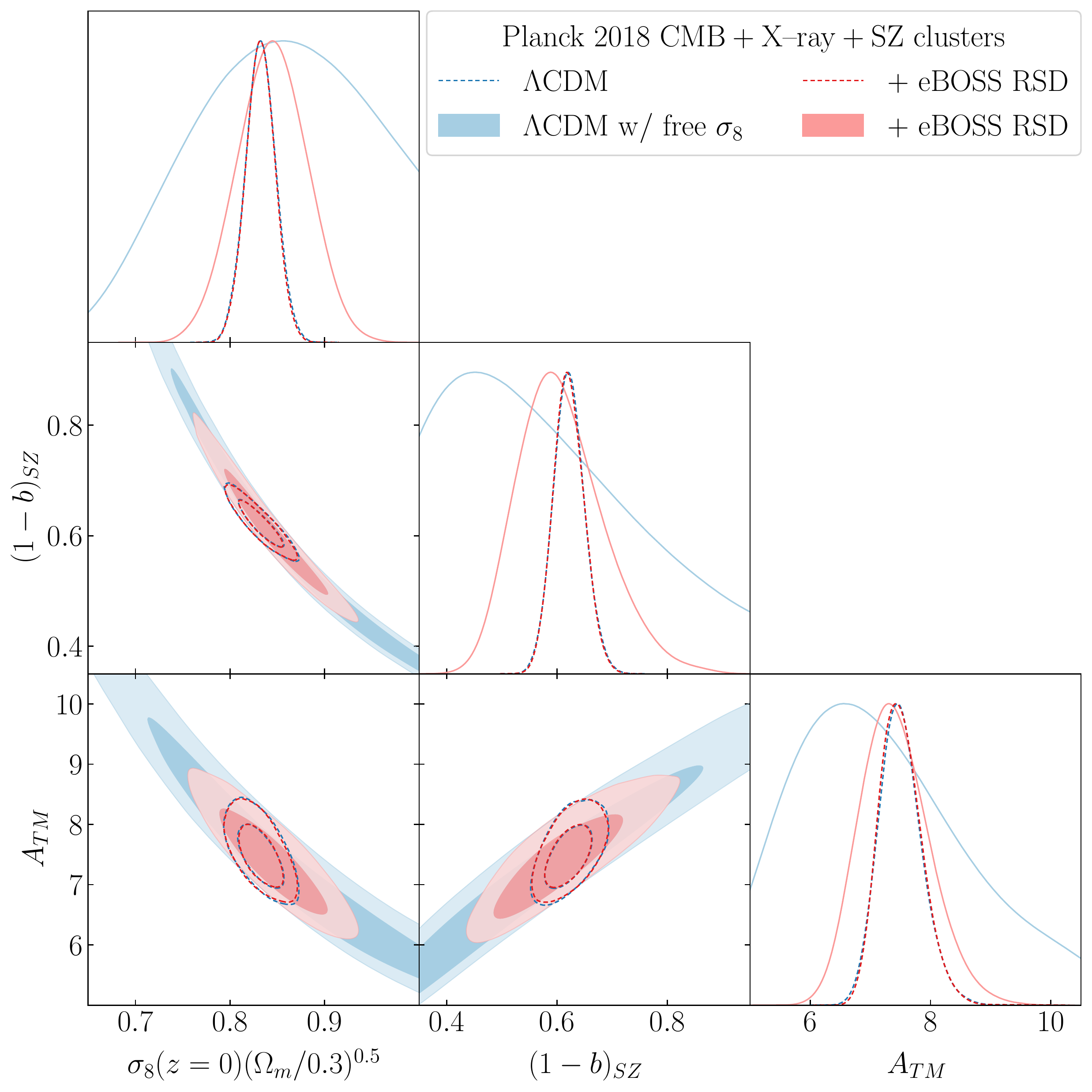}
  \caption{Confidence contours (68 and 95\%) and marginalised posterior distributions for parameters $S_8 = \sigma_8(\Omega_m/0.3)^{0.5}$, X-ray cluster mass calibration $A_{TM}$, and SZ cluster mass calibration $(1-b)$, with different combinations of cluster data (for the D16 mass function), eBOSS RSD data, and Planck CMB data. The filled contours and solid lines} show results for a $\Lambda$CDM model where the low-redshift amplitude of matter fluctuations is left as a free parameter; i.e. relying on the self-calibration approach. In blue, only cluster data are used as low-redshift probes, whereas we further add the eBOSS RSD data for the red contours.
  The dashed contours and lines are the corresponding constraints obtained when $\sigma_8$ is derived as a $\Lambda$CDM prediction. The effect of adding the eBOSS RSD data is imperceptible (blue dashed versus red dashed).
  \label{Triplot_S8}
\end{figure}

The above self-calibration approach amounts to inferring the late $\sigma_8$ from the eBOSS data using priors from Planck on all other cosmological parameters (and thus only on the shape of the power spectrum of matter fluctuations but not its amplitude). Our final constraints on $\sigma_8$  therefore essentially results from the local RSD measurements with priors on other cosmological parameters obtained from Planck.
This allows us to obtain joint constraints on the late amplitude of matter fluctuations and the calibration of cluster scaling relations.
Our final constraints are fully summarised in Figure~\ref{Triplot_S8}, where confidence (68\% and 95\%) contours and 1D marginalised posteriors are shown for our parameters of interest, namely the cluster calibration parameters, as well as the standard $S_8$ parameter:
\begin{equation}
    S_8 \equiv \sigma_8 \left(\frac{\Omega_m}{0.3}\right)^{0.5} \,
,\end{equation}
 to quantify the amplitude of matter fluctuations. To obtain those constraints, we performed a Markov chain Monte Carlo analysis using the ECLAIR \citep{2021PhRvD.104d3520I} and CosmoMC \citep{2002PhRvD..66j3511L,2013PhRvD..87j3529L} suites of codes, varying the full range of standard cosmological and nuisance parameters associated with the data sets used.

To practically implement the approach explained earlier, 
we considered a $\Lambda$CDM model where the overall amplitude of matter fluctuations $\sigma_8$ is freed in the range of late redshifts ($z\sim[0,1]$).
This approach was previously adopted by \citet{2019A&A...631A..96I}. As anticipated, it does not lead to stringent constraints on cluster calibrations when combined only with CMB data (cf.~blue solid lines and filled contours in Fig.~\ref{Triplot_S8}).

The SZ cluster sample is indeed too small to allow a stringent self-calibration, even if the background parameters are restricted by the Planck 2018 CMB constraints. We therefore repeated the same self-calibration approach, but added the eBOSS RSD data to the CMB and cluster counts. This new approach allows a combination of local cosmological data with Planck CMB constraints on cosmological parameters but without adding any prior on the amplitude of fluctuations, and thereby not on the calibration either, which can be now estimated from the combined set of data. We obtained tight constraints (68\% confidence limits thereafter, unless stated otherwise) on the corresponding parameters:
\begin{equation}
(1-b)= 0.608^{+0.063}_{-0.089} \: \rm{\: and \:} \: A_{TM} = 7.48^{+0.55}_{-0.68}
,\end{equation}
as illustrated by the solid red lines and filled contours in Fig.~\ref{Triplot_S8}. These low-redshift-based constraints on the calibration constitute the main conclusion of our analysis: in the context of the $\Lambda$CDM model, our calibration derived from the low-redshift eBOSS data is 3\,$\sigma$ away from its $(1-b) \sim 0.8$ value. We also produced tight constraints on the $S_8$ parameter:
\begin{equation}
    S_8 \equiv \sigma_8 \left(\frac{\Omega_m}{0.3}\right)^{0.5} = 0.841\pm 0.038 \,.
\end{equation}
These values are entirely consistent with those obtained in the standard analysis of the $\Lambda$CDM model (i.e.~where $\sigma_8$ is a derived parameter), for which $(1-b)= 0.620 \pm 0.029$ and $S_8= 0.834 \pm 0.016$ when using the combination of clusters and the full Planck CMB data (dashed blue). The addition of RSD data from eBOSS does not lead to any appreciable improvement (dashed red). This calibration is also consistent with weak lensing mass estimates from \cite{2014MNRAS.439....2V}, but lies at the extremes of various observational direct determinations. Some statistical analyses based on clusters also lead to results consistent with our conclusions \citep{2014MNRAS.439.1628Z}. Finally, similarly to \citet{2019A&A...631A..96I}, we examined the constraints obtained when extending the $\Lambda$CDM with a modification of gravity via the $\gamma$ parametrisation. The use of the Planck 2018 CMB data leads to
\begin{equation}
    (1-b) = 0.652^{+0.054}_{-0.062}.
\end{equation}
This value is 1\,$\sigma$ higher than when Planck 2015 is used \citep{2019A&A...631A..96I} and is not significantly improved by the addition of the eBOSS RSD data.

The SZ signal from clusters also contributes to the average fluctuations in the CMB \citep{1988MNRAS.233..637C} and also induces a global $y$ Compton distortion \citep{1996A&A...314...13B}. Planck mapped the tSZ component over the sky accurately, and the power spectrum of this component can be used as a useful source of constraints on cosmological parameters \citep{2002MNRAS.336.1256K,2018A&A...614A..13S}. Several recent works have applied this approach in combination with weak lensing surveys or galaxy surveys to obtain constraints on the calibration \citep{2020MNRAS.491.5464K,2020ApJ...902...56C}. A direct comparison with our approach is, however, difficult as the ingredients are different.
Our analysis for $\Lambda$CDM models, which leads to a calibration tightly constrained resulting from the addition of the eBOSS low-redshift data, sheds new light on the calibration issue.

\section{Conclusion}
\label{sec:Conclusion}

In the present work, we addressed the issue of the calibrations used in cluster mass--observable relations in a self-consistent way in the context of the $\Lambda$CDM model. For this purpose, we derived posterior distributions on calibration parameters, using the latest Planck CMB data, SZ and X-ray clusters counts, as well as SDSS RSD data. We left the amplitude of matter fluctuations at late times as a free parameter; that is, independent of the early-Universe constraints of CMB data.

As such, we obtained calibrations
relying essentially on the low-redshift data from eBOSS to constrain the late amplitude of matter fluctuations, with priors on other cosmological parameters obtained from Planck. This provides the first stringent constraints from the self-calibration approach obtained 
 via low-redshift data. 
 The resulting constraints on the calibrations $(1-b)$ and $A_{TM}$ for $\Lambda$CDM are entirely consistent with the values  obtained in the standard analysis in which the amplitude is derived directly from the Planck CMB data.
The preferred value of the SZ mass calibration parameter, $(1-b)$ remains on the order of 0.6 and 3\,$\sigma$ away from the fiducial value adopted in \cite{2014A&A...571A..20P}.
These calibrations stem from the amplitude of matter fluctuations $\sigma_8$ -- or equivalently the $S_8$ parameter -- entirely consistent with those derived in the standard analysis of the $\Lambda$CDM case from the full Planck data.
 It is worth noting that the recent DES Y3 results also report an amplitude $\sigma_8$ for $\Lambda$CDM models consistent with Planck 2018, while their cosmic-shear-only analysis may suffer from theoretical uncertainties in their modelling \citep{2021arXiv210513543A}. After the first version of this paper was submitted, the DES collaboration presented their analysis of the cross-correlation signal between their lensing map and the thermal SZ effect from Planck and ACT, allowing a determination of the calibration $(1-b) = 0.56 \pm 0.02$ \citep{2021arXiv210801601P}. Those results seem to lean towards conclusions similar to ours.

Fixing the cluster mass cluster calibrations is a critical ingredient for cluster cosmology and for the understanding of gas physics in clusters: the role of non-gravitational heating processes is known to be decisive in determining their final observational properties. Clearly, if a value of the $(1-b)$ calibration close to 0.6 can be consolidated, it would point towards consistency between the present day amplitude of matter fluctuations as measured by late-Universe probes and the one inferred from CMB data in the $\Lambda$CDM model. This would consequently call for 
a revision of the current models of non-gravitational physics in clusters shaping their baryonic component. Additionally, we expect a significant improvement of the determination of clusters masses in the coming
decade, thanks to the advent of the next generation of large-scale surveys such as Euclid\footnote{\url{http://www.euclid-ec.org/}}, the Vera C. Rubin Observatory Legacy Survey of Space and Time (LSST)\footnote{\url{https://www.lsst.org}} , or the Nancy Grace Roman Space Telescope\footnote{\url{https://wfirst.gsfc.nasa.gov}}. Not only will such surveys greatly increase the overall statistics of detected clusters, but unprecedented weak lensing measurements should shed additional light on those objects and provide robust and independent estimates of cluster masses.

\bibliographystyle{aa}
\bibliography{References}

\begin{thebibliography}{53}
\expandafter\ifx\csname natexlab\endcsname\relax\def\natexlab#1{#1}\fi

\bibitem[{{Abbott} {et~al.}(2018){Abbott}, {Abdalla}, {Alarcon}, {Aleksi{\'c}},
  {Allam}, {Allen}, {Amara}, {Annis}, {Asorey}, {Avila}, {Bacon}, {Balbinot},
  {Banerji}, {Banik}, {Barkhouse}, {Baumer}, {Baxter}, {Bechtol}, {Becker},
  {Benoit-L{\'e}vy}, {Benson}, {Bernstein}, {Bertin}, {Blazek}, {Bridle},
  {Brooks}, {Brout}, {Buckley-Geer}, {Burke}, {Busha}, {Campos}, {Capozzi},
  {Carnero Rosell}, {Carrasco Kind}, {Carretero}, {Castander}, {Cawthon},
  {Chang}, {Chen}, {Childress}, {Choi}, {Conselice}, {Crittenden}, {Crocce},
  {Cunha}, {D'Andrea}, {da Costa}, {Das}, {Davis}, {Davis}, {De Vicente},
  {DePoy}, {DeRose}, {Desai}, {Diehl}, {Dietrich}, {Dodelson}, {Doel},
  {Drlica-Wagner}, {Eifler}, {Elliott}, {Elsner}, {Elvin-Poole}, {Estrada},
  {Evrard}, {Fang}, {Fernandez}, {Fert{\'e}}, {Finley}, {Flaugher}, {Fosalba},
  {Friedrich}, {Frieman}, {Garc{\'\i}a-Bellido}, {Garcia-Fernandez}, {Gatti},
  {Gaztanaga}, {Gerdes}, {Giannantonio}, {Gill}, {Glazebrook}, {Goldstein},
  {Gruen}, {Gruendl}, {Gschwend}, {Gutierrez}, {Hamilton}, {Hartley}, {Hinton},
  {Honscheid}, {Hoyle}, {Huterer}, {Jain}, {James}, {Jarvis}, {Jeltema},
  {Johnson}, {Johnson}, {Kacprzak}, {Kent}, {Kim}, {King}, {Kirk}, {Kokron},
  {Kovacs}, {Krause}, {Krawiec}, {Kremin}, {Kuehn}, {Kuhlmann}, {Kuropatkin},
  {Lacasa}, {Lahav}, {Li}, {Liddle}, {Lidman}, {Lima}, {Lin}, {MacCrann},
  {Maia}, {Makler}, {Manera}, {March}, {Marshall}, {Martini}, {McMahon},
  {Melchior}, {Menanteau}, {Miquel}, {Miranda}, {Mudd}, {Muir}, {M{\"o}ller},
  {Neilsen}, {Nichol}, {Nord}, {Nugent}, {Ogando}, {Palmese}, {Peacock},
  {Peiris}, {Peoples}, {Percival}, {Petravick}, {Plazas}, {Porredon}, {Prat},
  {Pujol}, {Rau}, {Refregier}, {Ricker}, {Roe}, {Rollins}, {Romer}, {Roodman},
  {Rosenfeld}, {Ross}, {Rozo}, {Rykoff}, {Sako}, {Salvador}, {Samuroff},
  {S{\'a}nchez}, {Sanchez}, {Santiago}, {Scarpine}, {Schindler}, {Scolnic},
  {Secco}, {Serrano}, {Sevilla-Noarbe}, {Sheldon}, {Smith}, {Smith}, {Smith},
  {Soares-Santos}, {Sobreira}, {Suchyta}, {Tarle}, {Thomas}, {Troxel},
  {Tucker}, {Tucker}, {Uddin}, {Varga}, {Vielzeuf}, {Vikram}, {Vivas},
  {Walker}, {Wang}, {Wechsler}, {Weller}, {Wester}, {Wolf}, {Yanny}, {Yuan},
  {Zenteno}, {Zhang}, {Zhang}, {Zuntz}, \& {Dark Energy Survey
  Collaboration}}]{2018PhRvD..98d3526A}
{Abbott}, T.~M.~C., {Abdalla}, F.~B., {Alarcon}, A., {et~al.} 2018, \prd, 98,
  043526

\bibitem[{{Abbott} {et~al.}(2020){Abbott}, {Aguena}, {Alarcon}, {Allam},
  {Allen}, {Annis}, {Avila}, {Bacon}, {Bechtol}, {Bermeo}, {Bernstein},
  {Bertin}, {Bhargava}, {Bocquet}, {Brooks}, {Brout}, {Buckley-Geer}, {Burke},
  {Carnero Rosell}, {Carrasco Kind}, {Carretero}, {Castander}, {Cawthon},
  {Chang}, {Chen}, {Choi}, {Costanzi}, {Crocce}, {da Costa}, {Davis}, {De
  Vicente}, {DeRose}, {Desai}, {Diehl}, {Dietrich}, {Dodelson}, {Doel},
  {Drlica-Wagner}, {Eckert}, {Eifler}, {Elvin-Poole}, {Estrada}, {Everett},
  {Evrard}, {Farahi}, {Ferrero}, {Flaugher}, {Fosalba}, {Frieman},
  {Garc{\'\i}a-Bellido}, {Gatti}, {Gaztanaga}, {Gerdes}, {Giannantonio},
  {Giles}, {Grandis}, {Gruen}, {Gruendl}, {Gschwend}, {Gutierrez}, {Hartley},
  {Hinton}, {Hollowood}, {Honscheid}, {Hoyle}, {Huterer}, {James}, {Jarvis},
  {Jeltema}, {Johnson}, {Johnson}, {Kent}, {Krause}, {Kron}, {Kuehn},
  {Kuropatkin}, {Lahav}, {Li}, {Lidman}, {Lima}, {Lin}, {MacCrann}, {Maia},
  {Mantz}, {Marshall}, {Martini}, {Mayers}, {Melchior}, {Mena-Fern{\'a}ndez},
  {Menanteau}, {Miquel}, {Mohr}, {Nichol}, {Nord}, {Ogando}, {Palmese},
  {Paz-Chinch{\'o}n}, {Plazas}, {Prat}, {Rau}, {Romer}, {Roodman}, {Rooney},
  {Rozo}, {Rykoff}, {Sako}, {Samuroff}, {S{\'a}nchez}, {Sanchez}, {Saro},
  {Scarpine}, {Schubnell}, {Scolnic}, {Serrano}, {Sevilla-Noarbe}, {Sheldon},
  {Smith}, {Smith}, {Suchyta}, {Swanson}, {Tarle}, {Thomas}, {To}, {Troxel},
  {Tucker}, {Varga}, {von der Linden}, {Walker}, {Wechsler}, {Weller},
  {Wilkinson}, {Wu}, {Yanny}, {Zhang}, {Zhang}, {Zuntz}, \& {DES
  Collaboration}}]{2020PhRvD.102b3509A}
{Abbott}, T.~M.~C., {Aguena}, M., {Alarcon}, A., {et~al.} 2020, \prd, 102,
  023509

\bibitem[{{Aiola} {et~al.}(2020){Aiola}, {Calabrese}, {Maurin}, {Naess},
  {Schmitt}, {Abitbol}, {Addison}, {Ade}, {Alonso}, {Amiri}, {Amodeo},
  {Angile}, {Austermann}, {Baildon}, {Battaglia}, {Beall}, {Bean}, {Becker},
  {Bond}, {Bruno}, {Calafut}, {Campusano}, {Carrero}, {Chesmore}, {Cho},
  {Choi}, {Clark}, {Cothard}, {Crichton}, {Crowley}, {Darwish}, {Datta},
  {Denison}, {Devlin}, {Duell}, {Duff}, {Duivenvoorden}, {Dunkley},
  {D{\"u}nner}, {Essinger-Hileman}, {Fankhanel}, {Ferraro}, {Fox}, {Fuzia},
  {Gallardo}, {Gluscevic}, {Golec}, {Grace}, {Gralla}, {Guan}, {Hall},
  {Halpern}, {Han}, {Hargrave}, {Hasselfield}, {Helton}, {Henderson},
  {Hensley}, {Hill}, {Hilton}, {Hilton}, {Hincks}, {Hlo{\v{z}}ek}, {Ho},
  {Hubmayr}, {Huffenberger}, {Hughes}, {Infante}, {Irwin}, {Jackson}, {Klein},
  {Knowles}, {Koopman}, {Kosowsky}, {Lakey}, {Li}, {Li}, {Li}, {Lokken},
  {Louis}, {Lungu}, {MacInnis}, {Madhavacheril}, {Maldonado}, {Mallaby-Kay},
  {Marsden}, {McMahon}, {Menanteau}, {Moodley}, {Morton}, {Namikawa}, {Nati},
  {Newburgh}, {Nibarger}, {Nicola}, {Niemack}, {Nolta}, {Orlowski-Sherer},
  {Page}, {Pappas}, {Partridge}, {Phakathi}, {Pisano}, {Prince}, {Puddu}, {Qu},
  {Rivera}, {Robertson}, {Rojas}, {Salatino}, {Schaan}, {Schillaci}, {Sehgal},
  {Sherwin}, {Sierra}, {Sievers}, {Sifon}, {Sikhosana}, {Simon}, {Spergel},
  {Staggs}, {Stevens}, {Storer}, {Sunder}, {Switzer}, {Thorne}, {Thornton},
  {Trac}, {Treu}, {Tucker}, {Vale}, {Van Engelen}, {Van Lanen}, {Vavagiakis},
  {Wagoner}, {Wang}, {Ward}, {Wollack}, {Xu}, {Zago}, \&
  {Zhu}}]{2020JCAP...12..047A}
{Aiola}, S., {Calabrese}, E., {Maurin}, L., {et~al.} 2020, \jcap, 2020, 047

\bibitem[{{Alam} {et~al.}(2021){Alam}, {Aubert}, {Avila}, {Balland},
  {Bautista}, {Bershady}, {Bizyaev}, {Blanton}, {Bolton}, {Bovy}, {Brinkmann},
  {Brownstein}, {Burtin}, {Chabanier}, {Chapman}, {Choi}, {Chuang}, {Comparat},
  {Cousinou}, {Cuceu}, {Dawson}, {de la Torre}, {de Mattia}, {Agathe}, {des
  Bourboux}, {Escoffier}, {Etourneau}, {Farr}, {Font-Ribera}, {Frinchaboy},
  {Fromenteau}, {Gil-Mar{\'\i}n}, {Le Goff}, {Gonzalez-Morales},
  {Gonzalez-Perez}, {Grabowski}, {Guy}, {Hawken}, {Hou}, {Kong}, {Parker},
  {Klaene}, {Kneib}, {Lin}, {Long}, {Lyke}, {de la Macorra}, {Martini},
  {Masters}, {Mohammad}, {Moon}, {Mueller}, {Mu{\~n}oz-Guti{\'e}rrez}, {Myers},
  {Nadathur}, {Neveux}, {Newman}, {Noterdaeme}, {Oravetz}, {Oravetz},
  {Palanque-Delabrouille}, {Pan}, {Paviot}, {Percival}, {P{\'e}rez-R{\`a}fols},
  {Petitjean}, {Pieri}, {Prakash}, {Raichoor}, {Ravoux}, {Rezaie}, {Rich},
  {Ross}, {Rossi}, {Ruggeri}, {Ruhlmann-Kleider}, {S{\'a}nchez}, {S{\'a}nchez},
  {S{\'a}nchez-Gallego}, {Sayres}, {Schneider}, {Seo}, {Shafieloo}, {Slosar},
  {Smith}, {Stermer}, {Tamone}, {Tinker}, {Tojeiro}, {Vargas-Maga{\~n}a},
  {Variu}, {Wang}, {Weaver}, {Weijmans}, {Y{\`e}che}, {Zarrouk}, {Zhao},
  {Zhao}, \& {Zheng}}]{2021PhRvD.103h3533A}
{Alam}, S., {Aubert}, M., {Avila}, S., {et~al.} 2021, \prd, 103, 083533

\bibitem[{{Amon} {et~al.}(2021){Amon}, {Gruen}, {Troxel}, {MacCrann},
  {Dodelson}, {Choi}, {Doux}, {Secco}, {Samuroff}, {Krause}, {Cordero},
  {Myles}, {DeRose}, {Wechsler}, {Gatti}, {Navarro-Alsina}, {Bernstein},
  {Jain}, {Blazek}, {Alarcon}, {Fert{\'e}}, {Raveri}, {Lemos}, {Campos},
  {Prat}, {S{\'a}nchez}, {Jarvis}, {Alves}, {Andrade-Oliveira}, {Baxter},
  {Bechtol}, {Becker}, {Bridle}, {Camacho}, {Campos}, {Carnero Rosell},
  {Carrasco Kind}, {Cawthon}, {Chang}, {Chen}, {Chintalapati}, {Crocce},
  {Davis}, {Diehl}, {Drlica-Wagner}, {Eckert}, {Eifler}, {Elvin-Poole},
  {Everett}, {Fang}, {Fosalba}, {Friedrich}, {Giannini}, {Gruendl}, {Harrison},
  {Hartley}, {Herner}, {Huang}, {Huff}, {Huterer}, {Kuropatkin}, {Leget},
  {Liddle}, {McCullough}, {Muir}, {Pandey}, {Park}, {Porredon}, {Refregier},
  {Rollins}, {Roodman}, {Rosenfeld}, {Ross}, {Rykoff}, {Sanchez},
  {Sevilla-Noarbe}, {Sheldon}, {Shin}, {Troja}, {Tutusaus}, {Varga},
  {Weaverdyck}, {Yanny}, {Yin}, {Zhang}, {Zuntz}, {Aguena}, {Allam}, {Annis},
  {Bacon}, {Bertin}, {Bhargava}, {Brooks}, {Buckley-Geer}, {Burke},
  {Carretero}, {Costanzi}, {da Costa}, {Pereira}, {De Vicente}, {Desai},
  {Dietrich}, {Doel}, {Ferrero}, {Flaugher}, {Frieman}, {Garc{\'\i}a-Bellido},
  {Gaztanaga}, {Gerdes}, {Giannantonio}, {Gschwend}, {Gutierrez}, {Hinton},
  {Hollowood}, {Honscheid}, {Hoyle}, {James}, {Kron}, {Kuehn}, {Lahav}, {Lima},
  {Lin}, {Maia}, {Marshall}, {Martini}, {Melchior}, {Menanteau}, {Miquel},
  {Mohr}, {Morgan}, {Ogando}, {Palmese}, {Paz-Chinch{\'o}n}, {Petravick},
  {Pieres}, {Plazas Malag{\'o}n}, {Romer}, {Sanchez}, {Scarpine}, {Schubnell},
  {Serrano}, {Smith}, {Soares-Santos}, {Suchyta}, {Tarle}, {Thomas}, {To}, \&
  {Weller}}]{2021arXiv210513543A}
{Amon}, A., {Gruen}, D., {Troxel}, M.~A., {et~al.} 2021, arXiv e-prints,
  arXiv:2105.13543

\bibitem[{{Bahcall} \& {Cen}(1993)}]{1993ApJ...407L..49B}
{Bahcall}, N.~A. \& {Cen}, R. 1993, \apjl, 407, L49

\bibitem[{{Balland} \& {Blanchard}(1997)}]{1997ApJ...487...33B}
{Balland}, C. \& {Blanchard}, A. 1997, \apj, 487, 33

\bibitem[{{Barbosa} {et~al.}(1996){Barbosa}, {Bartlett}, {Blanchard}, \&
  {Oukbir}}]{1996A&A...314...13B}
{Barbosa}, D., {Bartlett}, J.~G., {Blanchard}, A., \& {Oukbir}, J. 1996, \aap,
  314, 13

\bibitem[{{Bartlett} \& {Silk}(1993)}]{1993ApJ...407L..45B}
{Bartlett}, J.~G. \& {Silk}, J. 1993, \apjl, 407, L45

\bibitem[{{Blanchard} \& {Bartlett}(1998)}]{bb}
{Blanchard}, A. \& {Bartlett}, J.~G. 1998, \aap, 332, L49

\bibitem[{{Blanchard} \& {Douspis}(2005)}]{2005A&A...436..411B}
{Blanchard}, A. \& {Douspis}, M. 2005, \aap, 436, 411

\bibitem[{{Bryan} \& {Norman}(1998)}]{1998ApJ...495...80B}
{Bryan}, G.~L. \& {Norman}, M.~L. 1998, \apj, 495, 80

\bibitem[{{Chiang} {et~al.}(2020){Chiang}, {Makiya}, {M{\'e}nard}, \&
  {Komatsu}}]{2020ApJ...902...56C}
{Chiang}, Y.-K., {Makiya}, R., {M{\'e}nard}, B., \& {Komatsu}, E. 2020, \apj,
  902, 56

\bibitem[{{Cole} \& {Kaiser}(1988)}]{1988MNRAS.233..637C}
{Cole}, S. \& {Kaiser}, N. 1988, \mnras, 233, 637

\bibitem[{{Delsart} {et~al.}(2010){Delsart}, {Barbosa}, \&
  {Blanchard}}]{2010A&A...524A..81D}
{Delsart}, P., {Barbosa}, D., \& {Blanchard}, A. 2010, \aap, 524, A81

\bibitem[{{DES Collaboration} {et~al.}(2021){DES Collaboration}, {Abbott},
  {Aguena}, {Alarcon}, {Allam}, {Alves}, {Amon}, {Andrade-Oliveira}, {Annis},
  {Avila}, {Bacon}, {Baxter}, {Bechtol}, {Becker}, {Bernstein}, {Bhargava},
  {Birrer}, {Blazek}, {Brandao-Souza}, {Bridle}, {Brooks}, {Buckley-Geer},
  {Burke}, {Camacho}, {Campos}, {Carnero Rosell}, {Carrasco Kind}, {Carretero},
  {Castander}, {Cawthon}, {Chang}, {Chen}, {Chen}, {Choi}, {Conselice},
  {Cordero}, {Costanzi}, {Crocce}, {da Costa}, {da Silva Pereira}, {Davis},
  {Davis}, {De Vicente}, {DeRose}, {Desai}, {Di Valentino}, {Diehl},
  {Dietrich}, {Dodelson}, {Doel}, {Doux}, {Drlica-Wagner}, {Eckert}, {Eifler},
  {Elsner}, {Elvin-Poole}, {Everett}, {Evrard}, {Fang}, {Farahi}, {Fernandez},
  {Ferrero}, {Fert{\'e}}, {Fosalba}, {Friedrich}, {Frieman},
  {Garc{\'\i}a-Bellido}, {Gatti}, {Gaztanaga}, {Gerdes}, {Giannantonio},
  {Giannini}, {Gruen}, {Gruendl}, {Gschwend}, {Gutierrez}, {Harrison},
  {Hartley}, {Herner}, {Hinton}, {Hollowood}, {Honscheid}, {Hoyle}, {Huff},
  {Huterer}, {Jain}, {James}, {Jarvis}, {Jeffrey}, {Jeltema}, {Kovacs},
  {Krause}, {Kron}, {Kuehn}, {Kuropatkin}, {Lahav}, {Leget}, {Lemos}, {Liddle},
  {Lidman}, {Lima}, {Lin}, {MacCrann}, {Maia}, {Marshall}, {Martini},
  {McCullough}, {Melchior}, {Mena-Fern{\'a}ndez}, {Menanteau}, {Miquel},
  {Mohr}, {Morgan}, {Muir}, {Myles}, {Nadathur}, {Navarro-Alsina}, {Nichol},
  {Ogando}, {Omori}, {Palmese}, {Pandey}, {Park}, {Paz-Chinch{\'o}n},
  {Petravick}, {Pieres}, {Plazas Malag{\'o}n}, {Porredon}, {Prat}, {Raveri},
  {Rodriguez-Monroy}, {Rollins}, {Romer}, {Roodman}, {Rosenfeld}, {Ross},
  {Rykoff}, {Samuroff}, {S{\'a}nchez}, {Sanchez}, {Sanchez}, {Sanchez Cid},
  {Scarpine}, {Schubnell}, {Scolnic}, {Secco}, {Serrano}, {Sevilla-Noarbe},
  {Sheldon}, {Shin}, {Smith}, {Soares-Santos}, {Suchyta}, {Swanson}, {Tabbutt},
  {Tarle}, {Thomas}, {To}, {Troja}, {Troxel}, {Tucker}, {Tutusaus}, {Varga},
  {Walker}, {Weaverdyck}, {Weller}, {Yanny}, {Yin}, {Zhang}, \&
  {Zuntz}}]{2021arXiv210513549D}
{DES Collaboration}, {Abbott}, T.~M.~C., {Aguena}, M., {et~al.} 2021, arXiv
  e-prints, arXiv:2105.13549

\bibitem[{{Despali} {et~al.}(2016){Despali}, {Giocoli}, {Angulo}, {Tormen},
  {Sheth}, {Baso}, \& {Moscardini}}]{2016MNRAS.456.2486D}
{Despali}, G., {Giocoli}, C., {Angulo}, R.~E., {et~al.} 2016, \mnras, 456, 2486

\bibitem[{{Euclid Collaboration} {et~al.}(2020){Euclid Collaboration},
  {Blanchard}, {Camera}, {Carbone}, {Cardone}, {Casas}, {Clesse}, {Ili{\'c}},
  {Kilbinger}, {Kitching}, {Kunz}, {Lacasa}, {Linder}, {Majerotto},
  {Markovi{\v{c}}}, {Martinelli}, {Pettorino}, {Pourtsidou}, {Sakr},
  {S{\'a}nchez}, {Sapone}, {Tutusaus}, {Yahia-Cherif}, {Yankelevich},
  {Andreon}, {Aussel}, {Balaguera-Antol{\'\i}nez}, {Baldi}, {Bardelli},
  {Bender}, {Biviano}, {Bonino}, {Boucaud}, {Bozzo}, {Branchini}, {Brau-Nogue},
  {Brescia}, {Brinchmann}, {Burigana}, {Cabanac}, {Capobianco}, {Cappi},
  {Carretero}, {Carvalho}, {Casas}, {Castander}, {Castellano}, {Cavuoti},
  {Cimatti}, {Cledassou}, {Colodro-Conde}, {Congedo}, {Conselice}, {Conversi},
  {Copin}, {Corcione}, {Coupon}, {Courtois}, {Cropper}, {Da Silva}, {de la
  Torre}, {Di Ferdinando}, {Dubath}, {Ducret}, {Duncan}, {Dupac}, {Dusini},
  {Fabbian}, {Fabricius}, {Farrens}, {Fosalba}, {Fotopoulou}, {Fourmanoit},
  {Frailis}, {Franceschi}, {Franzetti}, {Fumana}, {Galeotta}, {Gillard},
  {Gillis}, {Giocoli}, {G{\'o}mez-Alvarez}, {Graci{\'a}-Carpio}, {Grupp},
  {Guzzo}, {Hoekstra}, {Hormuth}, {Israel}, {Jahnke}, {Keihanen}, {Kermiche},
  {Kirkpatrick}, {Kohley}, {Kubik}, {Kurki-Suonio}, {Ligori}, {Lilje}, {Lloro},
  {Maino}, {Maiorano}, {Marggraf}, {Martinet}, {Marulli}, {Massey},
  {Medinaceli}, {Mei}, {Mellier}, {Metcalf}, {Metge}, {Meylan}, {Moresco},
  {Moscardini}, {Munari}, {Nichol}, {Niemi}, {Nucita}, {Padilla}, {Paltani},
  {Pasian}, {Percival}, {Pires}, {Polenta}, {Poncet}, {Pozzetti}, {Racca},
  {Raison}, {Renzi}, {Rhodes}, {Romelli}, {Roncarelli}, {Rossetti}, {Saglia},
  {Schneider}, {Scottez}, {Secroun}, {Sirri}, {Stanco}, {Starck}, {Sureau},
  {Tallada-Cresp{\'\i}}, {Tavagnacco}, {Taylor}, {Tenti}, {Tereno},
  {Toledo-Moreo}, {Torradeflot}, {Valenziano}, {Vassallo}, {Verdoes Kleijn},
  {Viel}, {Wang}, {Zacchei}, {Zoubian}, \& {Zucca}}]{IST:paper1}
{Euclid Collaboration}, {Blanchard}, A., {Camera}, S., {et~al.} 2020, \aap,
  642, A191

\bibitem[{{Hattori} \& {Matsuzawa}(1995)}]{1995A&A...300..637H}
{Hattori}, M. \& {Matsuzawa}, H. 1995, \aap, 300, 637

\bibitem[{{Henry}(1997)}]{1997ApJ...489L...1H}
{Henry}, J.~P. 1997, \apjl, 489, L1

\bibitem[{{Heymans} {et~al.}(2021){Heymans}, {Tr{\"o}ster}, {Asgari}, {Blake},
  {Hildebrandt}, {Joachimi}, {Kuijken}, {Lin}, {S{\'a}nchez}, {van den Busch},
  {Wright}, {Amon}, {Bilicki}, {de Jong}, {Crocce}, {Dvornik}, {Erben},
  {Fortuna}, {Getman}, {Giblin}, {Glazebrook}, {Hoekstra}, {Joudaki},
  {Kannawadi}, {K{\"o}hlinger}, {Lidman}, {Miller}, {Napolitano}, {Parkinson},
  {Schneider}, {Shan}, {Valentijn}, {Verdoes Kleijn}, \&
  {Wolf}}]{2021A&A...646A.140H}
{Heymans}, C., {Tr{\"o}ster}, T., {Asgari}, M., {et~al.} 2021, \aap, 646, A140

\bibitem[{{Hollinger} \& {Hudson}(2021)}]{2021MNRAS.tmp..238H}
{Hollinger}, A.~M. \& {Hudson}, M.~J. 2021, \mnras [\eprint[arXiv]{2101.04120}]

\bibitem[{{Hu}(2003)}]{Hu2003}
{Hu}, W. 2003, \prd, 67, 081304

\bibitem[{{Ili{\'c}} {et~al.}(2015){Ili{\'c}}, {Blanchard}, \&
  {Douspis}}]{2015A&A...582A..79I}
{Ili{\'c}}, S., {Blanchard}, A., \& {Douspis}, M. 2015, \aap, 582, A79

\bibitem[{{Ili{\'c}} {et~al.}(2021){Ili{\'c}}, {Kopp}, {Skordis}, \&
  {Thomas}}]{2021PhRvD.104d3520I}
{Ili{\'c}}, S., {Kopp}, M., {Skordis}, C., \& {Thomas}, D.~B. 2021, \prd, 104,
  043520

\bibitem[{{Ili{\'c}} {et~al.}(2019){Ili{\'c}}, {Sakr}, \&
  {Blanchard}}]{2019A&A...631A..96I}
{Ili{\'c}}, S., {Sakr}, Z., \& {Blanchard}, A. 2019, \aap, 631, A96

\bibitem[{{Jedamzik} {et~al.}(2021){Jedamzik}, {Pogosian}, \&
  {Zhao}}]{Jedamzik}
{Jedamzik}, K., {Pogosian}, L., \& {Zhao}, G.-B. 2021, Communications Physics,
  4, 123

\bibitem[{{Kaiser}(1986)}]{1986MNRAS.222..323K}
{Kaiser}, N. 1986, \mnras, 222, 323

\bibitem[{{Kaiser}(1991)}]{1991ApJ...383..104K}
{Kaiser}, N. 1991, \apj, 383, 104

\bibitem[{{Komatsu} \& {Seljak}(2002)}]{2002MNRAS.336.1256K}
{Komatsu}, E. \& {Seljak}, U. 2002, \mnras, 336, 1256

\bibitem[{{Koukoufilippas} {et~al.}(2020){Koukoufilippas}, {Alonso}, {Bilicki},
  \& {Peacock}}]{2020MNRAS.491.5464K}
{Koukoufilippas}, N., {Alonso}, D., {Bilicki}, M., \& {Peacock}, J.~A. 2020,
  \mnras, 491, 5464

\bibitem[{{Kravtsov} {et~al.}(2006){Kravtsov}, {Vikhlinin}, \&
  {Nagai}}]{2006ApJ...650..128K}
{Kravtsov}, A.~V., {Vikhlinin}, A., \& {Nagai}, D. 2006, \apj, 650, 128

\bibitem[{{Lemos} {et~al.}(2021){Lemos}, {Raveri}, {Campos}, {Park}, {Chang},
  {Weaverdyck}, {Huterer}, {Liddle}, {Blazek}, {Cawthon}, {Choi}, {DeRose},
  {Dodelson}, {Doux}, {Gatti}, {Gruen}, {Harrison}, {Krause}, {Lahav},
  {MacCrann}, {Muir}, {Prat}, {Rau}, {Rollins}, {Samuroff}, {Zuntz}, {Aguena},
  {Allam}, {Annis}, {Avila}, {Bacon}, {Bernstein}, {Bertin}, {Brooks}, {Burke},
  {Carnero Rosell}, {Carrasco Kind}, {Carretero}, {Castander}, {Conselice},
  {Costanzi}, {Crocce}, {Pereira}, {Davis}, {De Vicente}, {Desai}, {Diehl},
  {Doel}, {Eckert}, {Eifler}, {Elvin-Poole}, {Everett}, {Evrard}, {Ferrero},
  {Fert{\'e}}, {Flaugher}, {Fosalba}, {Frieman}, {Garc{\'\i}a-Bellido},
  {Gaztanaga}, {Gerdes}, {Giannantonio}, {Gruendl}, {Gschwend}, {Gutierrez},
  {Hartley}, {Hinton}, {Hollowood}, {Honscheid}, {Hoyle}, {Huff}, {James},
  {Jarvis}, {Lima}, {Maia}, {March}, {Marshall}, {Martini}, {Melchior},
  {Menanteau}, {Miquel}, {Mohr}, {Morgan}, {Myles}, {Ogando}, {Palmese},
  {Pandey}, {Paz-Chinch{\'o}n}, {Plazas Malag{\'o}n}, {Rodriguez-Monroy},
  {Roodman}, {Sanchez}, {Scarpine}, {Schubnell}, {Secco}, {Serrano},
  {Sevilla-Noarbe}, {Smith}, {Soares-Santos}, {Suchyta}, {Swanson}, {Tarle},
  {Thomas}, {To}, {Troxel}, {Varga}, {Weller}, {Wester}, {Wester}, \& {DES
  Collaboration}}]{2021MNRAS.505.6179L}
{Lemos}, P., {Raveri}, M., {Campos}, A., {et~al.} 2021, \mnras, 505, 6179

\bibitem[{{Lewis}(2013)}]{2013PhRvD..87j3529L}
{Lewis}, A. 2013, \prd, 87, 103529

\bibitem[{{Lewis} \& {Bridle}(2002)}]{2002PhRvD..66j3511L}
{Lewis}, A. \& {Bridle}, S. 2002, \prd, 66, 103511

\bibitem[{{Muir} {et~al.}(2021){Muir}, {Baxter}, {Miranda}, {Doux},
  {Fert{\'e}}, {Leonard}, {Huterer}, {Jain}, {Lemos}, {Raveri}, {Nadathur},
  {Campos}, {Chen}, {Dodelson}, {Elvin-Poole}, {Lee}, {Secco}, {Troxel},
  {Weaverdyck}, {Zuntz}, {Brout}, {Choi}, {Crocce}, {Davis}, {Gruen}, {Krause},
  {Lidman}, {MacCrann}, {M{\"o}ller}, {Prat}, {Ross}, {Sako}, {Samuroff},
  {S{\'a}nchez}, {Scolnic}, {Zhang}, {Abbott}, {Aguena}, {Allam}, {Annis},
  {Avila}, {Bacon}, {Bertin}, {Bhargava}, {Bridle}, {Brooks}, {Burke}, {Carnero
  Rosell}, {Carrasco Kind}, {Carretero}, {Cawthon}, {Costanzi}, {da Costa},
  {Pereira}, {Desai}, {Diehl}, {Dietrich}, {Doel}, {Estrada}, {Everett},
  {Evrard}, {Ferrero}, {Flaugher}, {Frieman}, {Garc{\'\i}a-Bellido},
  {Giannantonio}, {Gruendl}, {Gschwend}, {Gutierrez}, {Hinton}, {Hollowood},
  {Honscheid}, {Hoyle}, {James}, {Jeltema}, {Kuehn}, {Kuropatkin}, {Lahav},
  {Lima}, {Maia}, {Menanteau}, {Miquel}, {Morgan}, {Myles}, {Palmese},
  {Paz-Chinch{\'o}n}, {Plazas}, {Romer}, {Roodman}, {Sanchez}, {Scarpine},
  {Serrano}, {Sevilla-Noarbe}, {Smith}, {Suchyta}, {Swanson}, {Tarle},
  {Thomas}, {To}, {Tucker}, {Varga}, {Weller}, {Wilkinson}, \& {DES
  Collaboration}}]{2021PhRvD.103b3528M}
{Muir}, J., {Baxter}, E., {Miranda}, V., {et~al.} 2021, \prd, 103, 023528

\bibitem[{{Oukbir} \& {Blanchard}(1992)}]{1992A&A...262L..21O}
{Oukbir}, J. \& {Blanchard}, A. 1992, \aap, 262, L21

\bibitem[{{Pan} {et~al.}(2018){Pan}, {Kaplinghat}, \&
  {Knox}}]{2018PhRvD..97j3531P}
{Pan}, Z., {Kaplinghat}, M., \& {Knox}, L. 2018, \prd, 97, 103531

\bibitem[{{Pandey} {et~al.}(2021){Pandey}, {Gatti}, {Baxter}, {Hill}, {Fang},
  {Doux}, {Giannini}, {Raveri}, {DeRose}, {Huang}, {Moser}, {Battaglia},
  {Alarcon}, {Amon}, {Becker}, {Campos}, {Chang}, {Chen}, {Choi}, {Eckert},
  {Elvin-Poole}, {Everett}, {Ferte}, {Harrison}, {Maccrann}, {Mccullough},
  {Myles}, {Navarro Alsina}, {Prat}, {Rollins}, {Sanchez}, {Shin}, {Troxel},
  {Tutusaus}, {Yin}, {Aguena}, {Allam}, {Andrade-Oliveira}, {Bernstein},
  {Bertin}, {Bolliet}, {Bond}, {Brooks}, {Calabrese}, {Carnero Rosell},
  {Carrasco Kind}, {Carretero}, {Cawthon}, {Costanzi}, {Crocce}, {da Costa},
  {Pereira}, {De Vicente}, {Desai}, {Diehl}, {Dietrich}, {Doel}, {Dunkley},
  {Everett}, {Evrard}, {Ferraro}, {Ferrero}, {Flaugher}, {Fosalba},
  {Garcia-Bellido}, {Gaztanaga}, {Gerdes}, {Giannantonio}, {Gruen}, {Gruendl},
  {Gschwend}, {Gutierrez}, {Herner}, {Hincks}, {Hinton}, {Hollowood},
  {Honscheid}, {Hughes}, {Huterer}, {Jain}, {James}, {Jeltema}, {Krause},
  {Kuehn}, {Lahav}, {Lima}, {Lokken}, {Madhavacheril}, {Maia}, {Mcmahon},
  {Melchior}, {Menanteau}, {Miquel}, {Mohr}, {Moodley}, {Morgan}, {Nati},
  {Niemack}, {Page}, {Palmese}, {Paz-Chinchon}, {Pieres}, {Plazas Malagon},
  {Rodriguez-Monroy}, {Romer}, {Sanchez}, {Scarpine}, {Schaan}, {Serrano},
  {Sevilla-Noarbe}, {Sheldon}, {Sherwin}, {Sifon}, {Smith}, {Soares-Santos},
  {Spergel}, {Suchyta}, {Swanson}, {Tarle}, {Thomas}, {To}, {Varga}, {Weller},
  {Wollack}, \& {Xu}}]{2021arXiv210801601P}
{Pandey}, S., {Gatti}, M., {Baxter}, E., {et~al.} 2021, arXiv e-prints,
  arXiv:2108.01601

\bibitem[{{Peebles}(1982)}]{Peebles82}
{Peebles}, P.~J.~E. 1982, \apjl, 263, L1

\bibitem[{{Peebles}(1984)}]{1984ApJ...284..439P}
{Peebles}, P.~J.~E. 1984, \apj, 284, 439

\bibitem[{{Pierpaoli} {et~al.}(2003){Pierpaoli}, {Borgani}, {Scott}, \&
  {White}}]{2003MNRAS.342..163P}
{Pierpaoli}, E., {Borgani}, S., {Scott}, D., \& {White}, M. 2003, \mnras, 342,
  163

\bibitem[{{Planck Collaboration} {et~al.}(2014){Planck Collaboration}, {Ade},
  {Aghanim}, {Armitage-Caplan}, {Arnaud}, {Ashdown}, {Atrio-Barandela},
  {Aumont}, {Baccigalupi}, {Banday}, {Barreiro}, {Barrena}, {Bartlett},
  {Battaner}, {Battye}, {Benabed}, {Beno{\^\i}t}, {Benoit-L{\'e}vy}, {Bernard},
  {Bersanelli}, {Bielewicz}, {Bikmaev}, {Blanchard}, {Bobin}, {Bock},
  {B{\"o}hringer}, {Bonaldi}, {Bond}, {Borrill}, {Bouchet}, {Bourdin},
  {Bridges}, {Brown}, {Bucher}, {Burenin}, {Burigana}, {Butler}, {Cardoso},
  {Carvalho}, {Catalano}, {Challinor}, {Chamballu}, {Chary}, {Chiang},
  {Chiang}, {Chon}, {Christensen}, {Church}, {Clements}, {Colombi}, {Colombo},
  {Couchot}, {Coulais}, {Crill}, {Curto}, {Cuttaia}, {Da Silva}, {Dahle},
  {Danese}, {Davies}, {Davis}, {de Bernardis}, {de Rosa}, {de Zotti},
  {Delabrouille}, {Delouis}, {D{\'e}mocl{\`e}s}, {D{\'e}sert}, {Dickinson},
  {Diego}, {Dolag}, {Dole}, {Donzelli}, {Dor{\'e}}, {Douspis}, {Dupac},
  {Efstathiou}, {En{\ss}lin}, {Eriksen}, {Finelli}, {Flores-Cacho}, {Forni},
  {Frailis}, {Franceschi}, {Fromenteau}, {Galeotta}, {Ganga},
  {G{\'e}nova-Santos}, {Giard}, {Giardino}, {Giraud-H{\'e}raud},
  {Gonz{\'a}lez-Nuevo}, {G{\'o}rski}, {Gratton}, {Gregorio}, {Gruppuso},
  {Hansen}, {Hanson}, {Harrison}, {Henrot-Versill{\'e}},
  {Hern{\'a}ndez-Monteagudo}, {Herranz}, {Hildebrandt}, {Hivon}, {Hobson},
  {Holmes}, {Hornstrup}, {Hovest}, {Huffenberger}, {Hurier}, {Jaffe}, {Jaffe},
  {Jones}, {Juvela}, {Keih{\"a}nen}, {Keskitalo}, {Khamitov}, {Kisner},
  {Kneissl}, {Knoche}, {Knox}, {Kunz}, {Kurki-Suonio}, {Lagache},
  {L{\"a}hteenm{\"a}ki}, {Lamarre}, {Lasenby}, {Laureijs}, {Lawrence}, {Leahy},
  {Leonardi}, {Le{\'o}n-Tavares}, {Lesgourgues}, {Liddle}, {Liguori}, {Lilje},
  {Linden-V{\o}rnle}, {L{\'o}pez-Caniego}, {Lubin}, {Mac{\'\i}as-P{\'e}rez},
  {Maffei}, {Maino}, {Mandolesi}, {Marcos-Caballero}, {Maris}, {Marshall},
  {Martin}, {Mart{\'\i}nez-Gonz{\'a}lez}, {Masi}, {Matarrese}, {Matthai},
  {Mazzotta}, {Meinhold}, {Melchiorri}, {Melin}, {Mendes}, {Mennella},
  {Migliaccio}, {Mitra}, {Miville-Desch{\^e}nes}, {Moneti}, {Montier},
  {Morgante}, {Mortlock}, {Moss}, {Munshi}, {Naselsky}, {Nati}, {Natoli},
  {Netterfield}, {N{\o}rgaard-Nielsen}, {Noviello}, {Novikov}, {Novikov},
  {Osborne}, {Oxborrow}, {Paci}, {Pagano}, {Pajot}, {Paoletti}, {Partridge},
  {Pasian}, {Patanchon}, {Perdereau}, {Perotto}, {Perrotta}, {Piacentini},
  {Piat}, {Pierpaoli}, {Pietrobon}, {Plaszczynski}, {Pointecouteau}, {Polenta},
  {Ponthieu}, {Popa}, {Poutanen}, {Pratt}, {Pr{\'e}zeau}, {Prunet}, {Puget},
  {Rachen}, {Rebolo}, {Reinecke}, {Remazeilles}, {Renault}, {Ricciardi},
  {Riller}, {Ristorcelli}, {Rocha}, {Roman}, {Rosset}, {Roudier},
  {Rowan-Robinson}, {Rubi{\~n}o-Mart{\'\i}n}, {Rusholme}, {Sandri}, {Santos},
  {Savini}, {Scott}, {Seiffert}, {Shellard}, {Spencer}, {Starck}, {Stolyarov},
  {Stompor}, {Sudiwala}, {Sunyaev}, {Sureau}, {Sutton}, {Suur-Uski}, {Sygnet},
  {Tauber}, {Tavagnacco}, {Terenzi}, {Toffolatti}, {Tomasi}, {Tristram},
  {Tucci}, {Tuovinen}, {T{\"u}rler}, {Umana}, {Valenziano}, {Valiviita}, {Van
  Tent}, {Vielva}, {Villa}, {Vittorio}, {Wade}, {Wandelt}, {Weller}, {White},
  {White}, {Yvon}, {Zacchei}, \& {Zonca}}]{2014A&A...571A..20P}
{Planck Collaboration}, {Ade}, P.~A.~R., {Aghanim}, N., {et~al.} 2014, \aap,
  571, A20

\bibitem[{{Planck Collaboration} {et~al.}(2016){Planck Collaboration}, {Ade},
  {Aghanim}, {Arnaud}, {Ashdown}, {Aumont}, {Baccigalupi}, {Banday},
  {Barreiro}, {Bartlett}, {Bartolo}, {Battaner}, {Battye}, {Benabed},
  {Beno{\^\i}t}, {Benoit-L{\'e}vy}, {Bernard}, {Bersanelli}, {Bielewicz},
  {Bock}, {Bonaldi}, {Bonavera}, {Bond}, {Borrill}, {Bouchet}, {Bucher},
  {Burigana}, {Butler}, {Calabrese}, {Cardoso}, {Catalano}, {Challinor},
  {Chamballu}, {Chary}, {Chiang}, {Christensen}, {Church}, {Clements},
  {Colombi}, {Colombo}, {Combet}, {Comis}, {Couchot}, {Coulais}, {Crill},
  {Curto}, {Cuttaia}, {Danese}, {Davies}, {Davis}, {de Bernardis}, {de Rosa},
  {de Zotti}, {Delabrouille}, {D{\'e}sert}, {Diego}, {Dolag}, {Dole},
  {Donzelli}, {Dor{\'e}}, {Douspis}, {Ducout}, {Dupac}, {Efstathiou}, {Elsner},
  {En{\ss}lin}, {Eriksen}, {Falgarone}, {Fergusson}, {Finelli}, {Forni},
  {Frailis}, {Fraisse}, {Franceschi}, {Frejsel}, {Galeotta}, {Galli}, {Ganga},
  {Giard}, {Giraud-H{\'e}raud}, {Gjerl{\o}w}, {Gonz{\'a}lez-Nuevo},
  {G{\'o}rski}, {Gratton}, {Gregorio}, {Gruppuso}, {Gudmundsson}, {Hansen},
  {Hanson}, {Harrison}, {Henrot-Versill{\'e}}, {Hern{\'a}ndez-Monteagudo},
  {Herranz}, {Hildebrandt}, {Hivon}, {Hobson}, {Holmes}, {Hornstrup}, {Hovest},
  {Huffenberger}, {Hurier}, {Jaffe}, {Jaffe}, {Jones}, {Juvela},
  {Keih{\"a}nen}, {Keskitalo}, {Kisner}, {Kneissl}, {Knoche}, {Kunz},
  {Kurki-Suonio}, {Lagache}, {L{\"a}hteenm{\"a}ki}, {Lamarre}, {Lasenby},
  {Lattanzi}, {Lawrence}, {Leonardi}, {Lesgourgues}, {Levrier}, {Liguori},
  {Lilje}, {Linden-V{\o}rnle}, {L{\'o}pez-Caniego}, {Lubin},
  {Mac{\'\i}as-P{\'e}rez}, {Maggio}, {Maino}, {Mandolesi}, {Mangilli}, {Maris},
  {Martin}, {Mart{\'\i}nez-Gonz{\'a}lez}, {Masi}, {Matarrese}, {McGehee},
  {Meinhold}, {Melchiorri}, {Melin}, {Mendes}, {Mennella}, {Migliaccio},
  {Mitra}, {Miville-Desch{\^e}nes}, {Moneti}, {Montier}, {Morgante},
  {Mortlock}, {Moss}, {Munshi}, {Murphy}, {Naselsky}, {Nati}, {Natoli},
  {Netterfield}, {N{\o}rgaard-Nielsen}, {Noviello}, {Novikov}, {Novikov},
  {Oxborrow}, {Paci}, {Pagano}, {Pajot}, {Paoletti}, {Partridge}, {Pasian},
  {Patanchon}, {Pearson}, {Perdereau}, {Perotto}, {Perrotta}, {Pettorino},
  {Piacentini}, {Piat}, {Pierpaoli}, {Pietrobon}, {Plaszczynski},
  {Pointecouteau}, {Polenta}, {Popa}, {Pratt}, {Pr{\'e}zeau}, {Prunet},
  {Puget}, {Rachen}, {Rebolo}, {Reinecke}, {Remazeilles}, {Renault}, {Renzi},
  {Ristorcelli}, {Rocha}, {Roman}, {Rosset}, {Rossetti}, {Roudier},
  {Rubi{\~n}o-Mart{\'\i}n}, {Rusholme}, {Sandri}, {Santos}, {Savelainen},
  {Savini}, {Scott}, {Seiffert}, {Shellard}, {Spencer}, {Stolyarov}, {Stompor},
  {Sudiwala}, {Sunyaev}, {Sutton}, {Suur-Uski}, {Sygnet}, {Tauber}, {Terenzi},
  {Toffolatti}, {Tomasi}, {Tristram}, {Tucci}, {Tuovinen}, {T{\"u}rler},
  {Umana}, {Valenziano}, {Valiviita}, {Van Tent}, {Vielva}, {Villa}, {Wade},
  {Wandelt}, {Wehus}, {Weller}, {White}, {Yvon}, {Zacchei}, \&
  {Zonca}}]{2016A&A...594A..24P}
{Planck Collaboration}, {Ade}, P.~A.~R., {Aghanim}, N., {et~al.} 2016, \aap,
  594, A24

\bibitem[{{Planck Collaboration} {et~al.}(2020{\natexlab{a}}){Planck
  Collaboration}, {Aghanim}, {Akrami}, {Ashdown}, {Aumont}, {Baccigalupi},
  {Ballardini}, {Banday}, {Barreiro}, {Bartolo}, {Basak}, {Battye}, {Benabed},
  {Bernard}, {Bersanelli}, {Bielewicz}, {Bock}, {Bond}, {Borrill}, {Bouchet},
  {Boulanger}, {Bucher}, {Burigana}, {Butler}, {Calabrese}, {Cardoso},
  {Carron}, {Challinor}, {Chiang}, {Chluba}, {Colombo}, {Combet}, {Contreras},
  {Crill}, {Cuttaia}, {de Bernardis}, {de Zotti}, {Delabrouille}, {Delouis},
  {Di Valentino}, {Diego}, {Dor{\'e}}, {Douspis}, {Ducout}, {Dupac}, {Dusini},
  {Efstathiou}, {Elsner}, {En{\ss}lin}, {Eriksen}, {Fantaye}, {Farhang},
  {Fergusson}, {Fernandez-Cobos}, {Finelli}, {Forastieri}, {Frailis},
  {Fraisse}, {Franceschi}, {Frolov}, {Galeotta}, {Galli}, {Ganga},
  {G{\'e}nova-Santos}, {Gerbino}, {Ghosh}, {Gonz{\'a}lez-Nuevo}, {G{\'o}rski},
  {Gratton}, {Gruppuso}, {Gudmundsson}, {Hamann}, {Handley}, {Hansen},
  {Herranz}, {Hildebrandt}, {Hivon}, {Huang}, {Jaffe}, {Jones}, {Karakci},
  {Keih{\"a}nen}, {Keskitalo}, {Kiiveri}, {Kim}, {Kisner}, {Knox},
  {Krachmalnicoff}, {Kunz}, {Kurki-Suonio}, {Lagache}, {Lamarre}, {Lasenby},
  {Lattanzi}, {Lawrence}, {Le Jeune}, {Lemos}, {Lesgourgues}, {Levrier},
  {Lewis}, {Liguori}, {Lilje}, {Lilley}, {Lindholm}, {L{\'o}pez-Caniego},
  {Lubin}, {Ma}, {Mac{\'\i}as-P{\'e}rez}, {Maggio}, {Maino}, {Mandolesi},
  {Mangilli}, {Marcos-Caballero}, {Maris}, {Martin}, {Martinelli},
  {Mart{\'\i}nez-Gonz{\'a}lez}, {Matarrese}, {Mauri}, {McEwen}, {Meinhold},
  {Melchiorri}, {Mennella}, {Migliaccio}, {Millea}, {Mitra},
  {Miville-Desch{\^e}nes}, {Molinari}, {Montier}, {Morgante}, {Moss}, {Natoli},
  {N{\o}rgaard-Nielsen}, {Pagano}, {Paoletti}, {Partridge}, {Patanchon},
  {Peiris}, {Perrotta}, {Pettorino}, {Piacentini}, {Polastri}, {Polenta},
  {Puget}, {Rachen}, {Reinecke}, {Remazeilles}, {Renzi}, {Rocha}, {Rosset},
  {Roudier}, {Rubi{\~n}o-Mart{\'\i}n}, {Ruiz-Granados}, {Salvati}, {Sandri},
  {Savelainen}, {Scott}, {Shellard}, {Sirignano}, {Sirri}, {Spencer},
  {Sunyaev}, {Suur-Uski}, {Tauber}, {Tavagnacco}, {Tenti}, {Toffolatti},
  {Tomasi}, {Trombetti}, {Valenziano}, {Valiviita}, {Van Tent}, {Vibert},
  {Vielva}, {Villa}, {Vittorio}, {Wandelt}, {Wehus}, {White}, {White},
  {Zacchei}, \& {Zonca}}]{Param-Planck2018}
{Planck Collaboration}, {Aghanim}, N., {Akrami}, Y., {et~al.}
  2020{\natexlab{a}}, \aap, 641, A6

\bibitem[{{Planck Collaboration} {et~al.}(2020{\natexlab{b}}){Planck
  Collaboration}, {Aghanim}, {Akrami}, {Ashdown}, {Aumont}, {Baccigalupi},
  {Ballardini}, {Banday}, {Barreiro}, {Bartolo}, {Basak}, {Benabed}, {Bernard},
  {Bersanelli}, {Bielewicz}, {Bock}, {Bond}, {Borrill}, {Bouchet}, {Boulanger},
  {Bucher}, {Burigana}, {Butler}, {Calabrese}, {Cardoso}, {Carron},
  {Casaponsa}, {Challinor}, {Chiang}, {Colombo}, {Combet}, {Crill}, {Cuttaia},
  {de Bernardis}, {de Rosa}, {de Zotti}, {Delabrouille}, {Delouis}, {Di
  Valentino}, {Diego}, {Dor{\'e}}, {Douspis}, {Ducout}, {Dupac}, {Dusini},
  {Efstathiou}, {Elsner}, {En{\ss}lin}, {Eriksen}, {Fantaye},
  {Fernandez-Cobos}, {Finelli}, {Frailis}, {Fraisse}, {Franceschi}, {Frolov},
  {Galeotta}, {Galli}, {Ganga}, {G{\'e}nova-Santos}, {Gerbino}, {Ghosh},
  {Giraud-H{\'e}raud}, {Gonz{\'a}lez-Nuevo}, {G{\'o}rski}, {Gratton},
  {Gruppuso}, {Gudmundsson}, {Hamann}, {Handley}, {Hansen}, {Herranz}, {Hivon},
  {Huang}, {Jaffe}, {Jones}, {Keih{\"a}nen}, {Keskitalo}, {Kiiveri}, {Kim},
  {Kisner}, {Krachmalnicoff}, {Kunz}, {Kurki-Suonio}, {Lagache}, {Lamarre},
  {Lasenby}, {Lattanzi}, {Lawrence}, {Le Jeune}, {Levrier}, {Lewis}, {Liguori},
  {Lilje}, {Lilley}, {Lindholm}, {L{\'o}pez-Caniego}, {Lubin}, {Ma},
  {Mac{\'\i}as-P{\'e}rez}, {Maggio}, {Maino}, {Mandolesi}, {Mangilli},
  {Marcos-Caballero}, {Maris}, {Martin}, {Mart{\'\i}nez-Gonz{\'a}lez},
  {Matarrese}, {Mauri}, {McEwen}, {Meinhold}, {Melchiorri}, {Mennella},
  {Migliaccio}, {Millea}, {Miville-Desch{\^e}nes}, {Molinari}, {Moneti},
  {Montier}, {Morgante}, {Moss}, {Natoli}, {N{\o}rgaard-Nielsen}, {Pagano},
  {Paoletti}, {Partridge}, {Patanchon}, {Peiris}, {Perrotta}, {Pettorino},
  {Piacentini}, {Polenta}, {Puget}, {Rachen}, {Reinecke}, {Remazeilles},
  {Renzi}, {Rocha}, {Rosset}, {Roudier}, {Rubi{\~n}o-Mart{\'\i}n},
  {Ruiz-Granados}, {Salvati}, {Sandri}, {Savelainen}, {Scott}, {Shellard},
  {Sirignano}, {Sirri}, {Spencer}, {Sunyaev}, {Suur-Uski}, {Tauber},
  {Tavagnacco}, {Tenti}, {Toffolatti}, {Tomasi}, {Trombetti}, {Valiviita}, {Van
  Tent}, {Vielva}, {Villa}, {Vittorio}, {Wandelt}, {Wehus}, {Zacchei}, \&
  {Zonca}}]{2020A&A...641A...5P}
{Planck Collaboration}, {Aghanim}, N., {Akrami}, Y., {et~al.}
  2020{\natexlab{b}}, \aap, 641, A5

\bibitem[{{Riess} {et~al.}(2021){Riess}, {Casertano}, {Yuan}, {Bowers},
  {Macri}, {Zinn}, \& {Scolnic}}]{Riess}
{Riess}, A.~G., {Casertano}, S., {Yuan}, W., {et~al.} 2021, \apjl, 908, L6

\bibitem[{{Salvati} {et~al.}(2018){Salvati}, {Douspis}, \&
  {Aghanim}}]{2018A&A...614A..13S}
{Salvati}, L., {Douspis}, M., \& {Aghanim}, N. 2018, \aap, 614, A13

\bibitem[{{Suginohara} \& {Ostriker}(1998)}]{1998ApJ...507...16S}
{Suginohara}, T. \& {Ostriker}, J.~P. 1998, \apj, 507, 16

\bibitem[{{Verde} {et~al.}(2019){Verde}, {Treu}, \& {Riess}}]{verde}
{Verde}, L., {Treu}, T., \& {Riess}, A.~G. 2019, Nature Astronomy, 3, 891

\bibitem[{{von der Linden} {et~al.}(2014){von der Linden}, {Allen},
  {Applegate}, {Kelly}, {Allen}, {Ebeling}, {Burchat}, {Burke}, {Donovan},
  {Morris}, {Blandford}, {Erben}, \& {Mantz}}]{2014MNRAS.439....2V}
{von der Linden}, A., {Allen}, M.~T., {Applegate}, D.~E., {et~al.} 2014,
  \mnras, 439, 2

\bibitem[{{White} {et~al.}(1993){White}, {Efstathiou}, \&
  {Frenk}}]{1993MNRAS.262.1023W}
{White}, S.~D.~M., {Efstathiou}, G., \& {Frenk}, C.~S. 1993, \mnras, 262, 1023

\bibitem[{{Zu} {et~al.}(2014){Zu}, {Weinberg}, {Rozo}, {Sheldon}, {Tinker}, \&
  {Becker}}]{2014MNRAS.439.1628Z}
{Zu}, Y., {Weinberg}, D.~H., {Rozo}, E., {et~al.} 2014, \mnras, 439, 1628

\end{thebibliography}

\end{document}